\documentclass[final,authoryear,3p,times,twocolumn]{elsarticle}
\usepackage{natbib}
\bibpunct{(}{)}{;}{a}{}{,}
\usepackage{amssymb,amsmath}
\usepackage{graphicx}
\usepackage{graphics}

\usepackage{epsfig}
\usepackage{color} 
\usepackage{txfonts}
\usepackage{float}

\usepackage{dcolumn}
\usepackage{bm}
\usepackage[english]{babel}
\usepackage{t1enc}
\usepackage{subfigure}
\usepackage{times}

\usepackage{threeparttable} 
\usepackage{lscape}

\journal{Astroparticle Physics}

\begin{document}

\begin{frontmatter}
\title{Ultra-High-Energy Cosmic Rays from Low-Luminosity Active Galactic Nuclei}
\author[iss,rcapa,mpifr]{Ioana Du\c{t}an\fnref{foot}}
\ead{idutan@spacescience.ro}
\author[iss,rcapa,mpifr]{Lauren\c{t}iu I. Caramete\fnref{foot}}
\ead{lcaramete@spacescience.ro}
\address[iss]{Institute of Space Science, Atomi\c{s}tilor 409, 077125 Bucharest-M\v{a}gurele, Romania}
\address[rcapa]{Research Center for Atomic Physics and Astrophysics, Atomi\c{s}tilor 405, 077125 Bucharest-M\v{a}gurele, Romania}
\address[mpifr]{Max Planck Institute for Radio Astronomy, Auf dem H\"{u}gel 69, 53121 Bonn, Germany}
\fntext[foot]{Member of the International Max Planck Research School (IMPRS) for Astronomy and Astrophysics at the Universities of Bonn and Cologne. Member of the Pierre Auger Collaboration.}
\date{\today}
\begin{abstract}We investigate the production of ultra-high-energy cosmic ray (UHECR) in relativistic jets from low-luminosity active galactic nuclei (LLAGN). We start by proposing a model for the UHECR contribution from the black holes (BHs) in LLAGN, which present a jet power $P_{\mathrm{j}} \leqslant 10^{46}$ erg s$^{-1}$. This is in contrast to the opinion that only high-luminosity AGN can accelerate particles to energies $ \geqslant 50$ EeV. We rewrite the equations which describe the synchrotron self-absorbed emission of a non-thermal particle distribution to obtain the observed radio flux density from sources with a flat-spectrum core and its relationship to the jet power. We find that the UHECR flux is dependent on the {\it observed radio flux density, the distance to the AGN, and the BH mass}, where the particle acceleration regions can be sustained by the magnetic energy extraction from the BH at the center of the AGN. We use a complete sample of 29 radio sources with a total flux density at 5 GHz 
greater than 0.5 Jy to make predictions for the maximum particle energy, luminosity, and flux of the UHECRs from nearby AGN. These predictions are then used in a semi-analytical code developed in Mathematica (SAM code) as inputs for the Monte-Carlo simulations to obtain the distribution of the arrival direction at the Earth and the energy spectrum of the UHECRs, taking into account their deflection in the intergalactic magnetic fields. For comparison, we also use the CRPropa code with the same initial conditions as for the SAM code. Importantly, to calculate the energy spectrum we also include the weighting of the UHECR flux per each UHECR source. Next, we compare the energy spectrum of the UHECRs with that obtained by the Pierre Auger Observatory.\end{abstract}
\end{frontmatter}
\parindent=0cm

\section{Introduction}

Cosmic rays (CRs) are a direct sample of matter from outside the solar system, and their study can, for instance, provide important information on the chemical evolution of the universe or improve constraints on Galactic and extragalactic magnetic fields. They can be measured indirectly through the study of extensive air showers that are induced as the CRs hit the top of the atmosphere (known as CR events). The extensive air showers are currently observed using air fluorescence [e.g., High Resolution Fly's Eye (HiRes) experiment\footnote{http://www.cosmic-ray.org}] or large array, ground-based detectors [e.g.,  Akeno Giant Air Shower Array (AGASA){\footnote{http://www-akeno.icrr.u-tokyo.ac.jp/AGASA}], or both [e.g., Pierre Auger Observatory (Auger)\footnote{http://www.auger.org}]. In the future, space-based detectors might be another option. UHECR particles are mostly protons or fully ionized nuclei with energy above 50 EeV (1 EeV = 10$^{18}$ eV). At such high energies, the flux of UHECRs is very low and 
only a few dozen particles per square kilometer per century are expected. This is one of the main reasons for the difficulty posed in understanding the origin and nature of the UHECRs. Therefore, very large detector arrays are required. The Pierre Auger Observatory, by far the biggest cosmic ray detection instrument, uses air fluorescence and water detection in a hybrid instrument with an aperture of 7000 km$^2$ sr.

Joint efforts have been made during the past decade by worldwide, cosmic ray experiments to help us understand from where the UHECRs come and what is their nature. It is believed that the UHECRs originate in extragalactic sources, as the gyroradius of a proton with an energy of 100~EeV is of the order of the dimension of our Galaxy, whereas most of the CR particles with energy below 50~EeV originate within our Galaxy \cite[e.g.,][]{berezinsky06,stanev10b,stanev10}. If the UHECR particles are protons, they are subject to energy loss by creating pions through their occasional collisions with the cosmic microwave background (CMB) photons. This process produces a suppression of the cosmic ray energy spectrum beyond 50 EeV, which is known as the Greisen-Zatsepin-Guzmin (GZK) cutoff \citep{greisen,zatsepin-kuzmin}. Therefore, the UHECRs would not be able to survive the propagation from their acceleration sites to us unless their sources are located within $\sim 100$~Mpc. The presence of the GZK cutoff at the 
expected energy in the data released by the HiRes collaboration was taken as strong evidence that the UHECR flux is dominated by protons \citep{data2}. 

A suppression of the CR flux has also been observed in the data released by the Pierre Auger collaboration \citep{spectrum,data1}. With respect to primary composition, this collaboration has exploited the observation of the longitudinal shower development with fluorescence detectors to measure the depth of the maximum of the shower evolution, $X_{\rm max}$, which is sensitive to the primary mass. A gradual increase of the average mass of cosmic rays with energy up to 59 EeV is deduced when comparing the absolute values of $X_{\rm max}$ and RMS($X_{\rm max}$) to air shower simulations \citep{data1b}.

The data collected by the Pierre Auger Observatory provide evidence for a correlation between the arrival directions of CR events above 55 EeV and the positions of AGN with $z < 0.018$ \citep{auger07,auger08,data1b,2011arXiv1107.4809T,auger13}, where the region around the position of the radio-galaxy Cen A has the largest excess of arrival directions relative to the isotropic expectations. The correlation is shown for a selection of AGN from the catalog of \citet{veron}, which do not necessarily follow the same structure as the gamma-ray bursts (GRBs). We emphasize that there is no clear detection of the UHECR sources, just a strong evidence for the anisotropy in the arrival directions of UHECRs.

At highest energies, heavy nuclei may be deflected by Galactic magnetic fields, whereas proton propagation is affected by the CMB, as well as by magnetic deflection, though to a less degree compared to that of particles with higher mass number \cite[e.g.,][]{medina98}.

UHECRs are most probably accelerated at astrophysical shocks, for instance, through a first-order Fermi mechanism \cite[e.g.,][]{ga99}, in very powerful systems that can be associated with jets and hot spots in AGN and GRBs \citep{waxman95,vietri95}, in large-scale shocks in clusters \cite[e.g.,][]{farrar06}, or as iron nuclei in pulsar winds from rapidly-spinning, young neutron stars \citep{blasi00,fang12}. Numerical simulations of particle acceleration in shocks have been widely performed using different values for the shock Lorentz factor and background conditions at the shock front \cite[e.g.,][]{bo98,acht,kirk00,kw05,jn06,meli}, which lead to a slope of the particle distribution of $p \sim [1.5,2.5]$.  Such shocks can also be associated with Poynting flux models for the origin of jets from force-free magnetosphere above accretion disks \cite[e.g.,][]{lovelace,bland76,boldt-ghosh,blandford00,biermann08}. Magnetic reconnection in relativistic jets represents another option for UHECR acceleration \
cite[e.g.,][]{giannios10}. As an alternative, \citet{Farrar09} showed that very intense, short-duration AGN flares that result from the tidal disruption of a star or from a disk instability can accelerate UHECRs. (See also \citet{waxman09}.)

In this paper, we propose a model for the UHECR contribution from relativistic jets in LLAGN and calculate the expected energy spectrum of UHECRs using the SAM code developed by \citet{biermann08}, \citet{laur11a}. For comparison, we also employ the CRPropa code, which is set up with the same initial conditions as the SAM code. The particles in the jet are powered by the BH accretion disk and then accelerated at relativistic shocks with energies up to the ultra-high energy (UHE) domain. We limit the launching area of the jets to the innermost part of the disk located inside the BH ergosphere, where the rotational effects of the space-time are very strong. There are several general relativistic magnetohydrodynamic (GRMHD) codes, the result of which show that the jets can be magnetically driven from a thin disk located inside the BH ergosphere via a Penrose-like process \citep{shinjiet99,ken05,mizuno07} or via the Blandford-Znajek mechanism (BZ, \citet{bz}) when a thick accretion disk is considered \citep{
mckinney06,mckinney09}. (But see 
the simulations by \citet{fragile12}, where the BZ driven jet does not depend on the disk thickness.) In contrast to the BZ mechanism, where the power of the jet is proportional to the square of the BH spin ($P_{\rm j} \sim a^2$), in the model presented here the dependence of $P_{\rm j}$ on $a$ comes through the launching area of the jets (see the equations in Appendices A and C). In the jets, the electrons lose their energy through synchrotron emission, whereas the protons, as well as heavy nuclei (here, iron nuclei), are capable of surviving the radiative cooling and, perhaps, of propagating through the intergalactic and Galactic medium towards us. Since the particle species undergo the same acceleration process, there must be a correlation between the electron synchrotron emission and the energy of the UHECR particles (protons and iron nuclei). We seek this correlation to make predictions for maximum energy, luminosity, and flux of the UHECRs from nearby LLAGN. This is in contrast to the 
opinion that only high-luminosity AGN can accelerate particles to UHE domain \cite[e.g.,][]{zaw09}.  Taking into account the deflection of the trajectories of the UHECRs in the intergalactic and Galactic magnetic fields, we calculate the distribution of the arrival direction at the Earth and the energy spectrum of the UHECRs. The latter is then compared with the energy spectrum obtained by the Pierre Auger Observatory. We point out that LLAGN as sources of UHECRs were also proposed by \citet{moskalenko09}, where discussions about the implication of AGN jet power and intergalactic, magnetic field configurations for the observed statistical correlation between AGN and UHECR events are presented. Our work is a step further to that of \citet{moskalenko09}, as we include quantitative estimations of UHECR flux using its correlation to the AGN jet power, as well as simulations of UHECR particle propagation in the intergalactic and Galactic magnetic fields. To obtain the energy spectrum of UHECRs, we use a 
complete sample of 29 LLAGN with a total radio flux density larger than 0.5 Jy \citep{biermann08,laur11a}. About 80\% of our sample is contained in the all-sky catalog of local radio galaxies of \citet{vanVelzen2012}, which is used to seek for the correlation between the UHECRs and LLAGN \citep{vanVelzen2013}. The fact that some sources of our sample are not included in the catalog by \citet{vanVelzen2012} might be attributed to the difference in the data; i.e., the frequency at which the radio flux density was measured: 5 GHz in our case and 1.5 GHz and 843 MHz in the case of \citet{vanVelzen2012}.

In Section \ref{model}, we provide a description of the model for the UHECR contribution from relativistic jets in LLAGN}. We derive the luminosity and flux of the UHECRs based on the relation between the jet power and the observed radio flux density for a flat-spectrum core source (see \ref{jetpower}) and calculate the particle maximum energy taking into account the spatial limit and synchrotron emission losses. In Section \ref{sources}, we provide the predictions for nearby galaxies as possible sources of UHECRs by employing the SAM code. For comparison, we also use the CRPropa code with the same input setup as for the SAM code. We then compare the results of the two codes with those obtained by the Pierre 
Auger Observatory. In Section \ref{sec:summary}, we present a summary of the key points and discuss the implication of this model for further studies of UHECRs.

\section{Description of the model for UHECR source}
\label{model} 

\subsection{Model conditions}

\begin{itemize}
\item We assume that the UHECRs are accelerated by shocks in AGN jets, which are launched from the inner accretion disk which is located inside the BH ergosphere, where the rotational effects of the space-time become much stronger \citep{eu04}. (The inner disk extends from the stationary limit $r_{\mathrm{sl}}$ inward to the innermost stable orbit $r_{\mathrm{ms}}$.) In the model by \citet{eu04,eu10}, the rotation of the space-time channels a fraction of the disk energy (i.e., the gravitational energy of the disk plus the rotational energy of the BH that is deposited into the disk by closed magnetic field lines, which connect the BH to the accretion disk) into a population of particles that escape from the disk surfaces, carrying away mass, energy, and angular momentum in the form of jets, allowing the remaining disk gas to accrete. The power of the jets can have two components; the one which comes from the accretion power and the other one which comes from the BH rotational energy. The latter 
component dominates when the mass accretion rate\footnote{Here, the term mass accretion rate refers to the mass flow rate through the disk up to the ergosphere ($\dot{M}_{\rm D} = \dot{m}\dot{M}_{\rm Edd}$), where $\dot{M}_{\rm Edd}$ is the Eddington accretion rate. $\dot{M}_{\rm Edd}$ is defined from the Eddington luminosity as ${\dot{M}}_{\mathrm{Edd}} = L_{\mathrm{Edd}}/(\varepsilon c^2) = 4\pi GM/(\varepsilon \kappa_{\mathrm{T}} c)$, where $\varepsilon$ is the efficiency of converting the gravitational energy of the accretion disk into radiation and $\kappa_{\mathrm{T}}$ denotes the Thomson opacity. $\varepsilon$ depends on the BH spin parameter as $\varepsilon=1-E^{\dagger}_{\mathrm{ms}}$ \citep{t74}, so that $\varepsilon= 0.06$ for a Schwarzschild BH and $\varepsilon= 0.42$ for an extremely spinning Kerr BH. If we scale the BH mass to $10^9 M_{\odot}$, then ${\dot{M}}_{\mathrm{Edd}} = {\dot{M}}_{\mathrm{Edd}}^{\dagger}\varepsilon^{-1}(M/10^9 M_{\odot})$, where ${\dot{M}}_{\mathrm{Edd}}^{\dagger} = 1.38\
times 10^{26}$ g s$^{-1}$.} relative to the Eddington accretion rate is below $\dot{m} \sim 10^{-1.8}$. 

\begin{figure}\centering
\epsfig{file=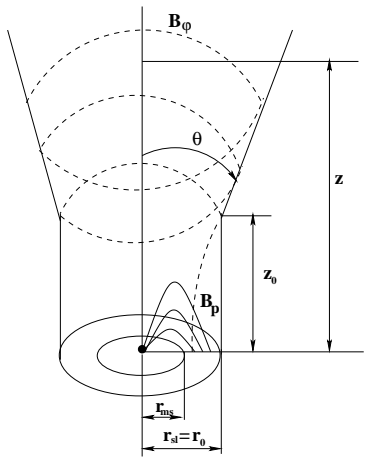,height=6cm}
\caption{Schematic representation of the jet geometry. The jet is launched from the inner disk, extending from the stationary limit inward to the innermost stable orbit, and propagates along a cylinder up to a distance of $z_{\mathrm{0}}$. Then, it expands freely into a conical geometry with a constant opening angle $2\,\theta$. The magnetic field lines threading the disk near the BH (dashed lines) are wound up, far from the BH, into a toroidal magnetic field $B^{\phi}$ that collimates the jet.}
\label{fig:jetGeom}  
\end{figure}

\item Being driven from the accretion disk, the jet propagates along a cylinder of length $z_{\mathrm{0}}$ (see Fig.~\ref{fig:jetGeom}), where the poloidal and toroidal components of the comoving magnetic field become approximately equal, and then extends into a conical shape with a constant opening angle $2\,\theta$, as a consequence of the free adiabatic expansion of the jet plasma (similar to \citet{markoff01}). The cylindrical surface represents the envelope of the magnetic field lines, which close to the BH are mainly parallel. (We do not include here a monopole-like configuration of the magnetic field.) We suppose that a shock is produced at the jet height $z = z_0$. (Production of a shock associated with the twisting of the magnetic field lines, where the toroidal component of the magnetic field dominates, was observed in GRMHD simulations. See, e.g., fig 3 in \citet{mizuno04}.)

\item As a result of the shock produced at $z_0$, a power-law energy distribution of the particles is established. The number density of the electrons in the energy interval $E$, $E + dE$  [or $m_{\mathrm{e}}c^2\gamma$, $m_{\mathrm{e}}c^2(\gamma + d\gamma$)] has, in terms of the Lorentz factor, the following power-law form: $N(\gamma) d\gamma=C' \gamma^{-p} d\gamma$. In the case of a conical jet, the normalization of the electron number density is \citep{bk79}: 
\begin{equation}
C'=C'_0\left( \frac{z}{z_0}\right)^{-2}  (\textrm{cm}^{-3}) .
\label{norm2}
\end{equation} 
We set the slope of the particle energy distribution to $p = 2.2, 2.3$, and $2.4$.
 
\item The calculations are performed for the case when the UHECRs are composed of 90\% protons and 10\% iron nuclei \cite[e.g.,][]{2008JCAP...10..033A}. 

\item We consider the strength of the BH magnetic field to have its maximum value. This condition provides, in turn, the minimum values of the particle maximum energy, luminosity, and flux of the UHECRs. 
\end{itemize}

\subsection{\label{magn}Magnetic field scaling along a steady jet}

To describe the jet physics, we use the following reference frames: (i) the frame comoving with the jet and (ii) the (rest) frame of the observer, in which the relativistic jet moves with the bulk Lorentz factor. In a frame comoving with the jet, the poloidal component of the magnetic field is considered to vary as $B_{\mathrm{p}} \sim z^{-2}$. This variation follows from the conservation of magnetic flux along the axis $z$. To keep the field divergence-free, the toroidal component must vary as $B_{\mathrm{\phi}} \sim z^{-1}$. This topology of $B_{\mathrm{\phi}} \sim z^{-1}$ was first derived by \citet{parker} for the magnetohydrodynamics solution of a spherical-symmetric flow (so that, a jet can be considered a conical cut along the flow surfaces). [See also \citet{bk79}.]
At the distance $z_0$, the poloidal and toroidal components of the comoving magnetic field become approximately equal $B_{\mathrm{p0}} \simeq B_{\mathrm{\phi0}}$. The distance $z_0$ might be a few tens of gravitational radii,\footnote{The gravitational radius is defined as $r_{\mathrm{g}} \equiv GM/c^2=r^{\dagger}_{\mathrm{g}}(M/10^9 M_{\odot})=1.48\times 10^{14}(M/10^9 M_{\odot})$ cm, where $G$ is the Newtonian gravitational constant, $M$ is the BH mass, and $c$ is the speed of light.} based on the fact that the VLBI observation, for instance, of the jet in M87 at 43 GHz gives evidence on the jet collimation (by the toroidal magnetic field) on scales of 60-200 $r_{\mathrm{g}}$ \citep{biretta02}. More recently, \citet{hada11} argued that 43 GHz VLBI core is located at $\sim 40\,r_{\mathrm{g}}$ from the central engine, where they performed the core shift measurement by using multi-frequency, phase-referencing Very Long Baseline Array observations, and that the measured frequency dependence of the core 
shift is in good agreement with a synchrotron self-absorbed jets. The structure of the M 87 jet, from sub-miliarsec to arcsec scales, was also investigated by \citet{nakamura13}, where the (bulk) acceleration and collimation of the jet are correlated in a parabolic streamline. The question is whether the streamline can be extrapolated continuously towards the jet launching region. To answer such question future observations with high sensitivity, e.g., a space VLBI and/or VLBI at higher frequency than 86 GHz are needed. Next, if we consider the $\gamma$-ray emission from M 87 and Cen A, the absorption of the TeV photons produced at about $\sim 100\,r_{\rm g}$ and $\sim 50\,r_{\rm g}$, respectively, from the BH is small, which means that the observed radiation could come from the BH vicinity  \citep{acciari09,brodatzki11,saba13}. Nevertheless, in calculating the luminosity and flux of the UHECRs we do not make use of the exact location of $z_0$, we rather extract the ratio of $z/z_0$ from the expression of 
the optical depth (Eq.~\ref{tau}). A large-scale and predominantly toroidal magnetic field can exert an inward force (hoop stress), confining and collimating the jet \cite[e.g.,][]{bisnovatyi-ruzmaikin,bp82}. The magnetic hoop stress is balanced either by the gas pressure of the jet or by centrifugal force if the jet is spinning. From $z_0 $ upward, the poloidal component of the magnetic field becomes weaker, so that the field lines are soon wound up in the azimuthal direction by the jet rotation. Thus, above $z_0$, the magnetic field along the jet is nearly azimuthal $B \sim B_{\phi}$ (for a steady jet) and varies inversely proportional to the distance along the jet:
\begin{equation}
B = B_0 \left( \frac{z}{z_0}\right)^{-1} , 
\label{magnfield2}
\end{equation}
where $B_0 \equiv B_{\mathrm{\phi0}} \simeq B_{\mathrm{p0}}$ is the strength of the magnetic field at the height $z=z_0$ above the equatorial plane of the BH. This $z$-dependence of the magnetic field appears to be contradicted by radio-polarization observations \citep{bridle&perley}. These observations strongly suggest that the magnetic field is predominantly parallel to the jet axis initially and only later becomes perpendicular to the jet axis, with some parallel magnetic field left over. \citet{pbb09} argued that the basic pattern of the magnetic field is indeed $B_{\phi} \sim z^{-1}$ and that the observational evidence for a parallel magnetic field is due to highly oblique shocks. Their argument is based on the observations of the jet structure which might be explained through the occurrence of the moving shocks between $\sim 10$ and $\sim 10^{3}$ $r_{\mathrm{g}}$ \citep{marscher}, while the first stationary, strong shock can be produced in the approximate range of $(3 - 6) \times10^{3}$ 
$r_{\mathrm{g}}$ \citep{markoff01,markoff05} or $\sim 10^{5}$ $r_{\mathrm{g}}$ as seen in blazars \citep{marscher}. 


The strength of the magnetic field in the comoving frame $B_0$ can be related to the poloidal magnetic field in the BH frame $B_{\mathrm{H}}$ \cite[e.g.,][]{drenkhahn} as
\begin{equation}
B_0=\frac{1}{\gamma_ {\mathrm{j}}}B_{\mathrm{H}}=\frac{B_{\mathrm{H}}^{\mathrm{max}}}{\gamma_ {\mathrm{j}}}\left(\frac{B_{\mathrm{H}}}{B_{\mathrm{H}}^{\mathrm{max}}} \right) ,
\label{field}
\end{equation} 
where the maximum value of the BH magnetic field is given by
\begin{equation}
B_{\mathrm{H}}^{\mathrm{max}} \simeq  0.56 \times 10^4 \left(\frac{M}{10^9M_{\odot}}\right) ^{-1/2}\ \mathrm{gauss},
\label{maxB}
\end{equation} 
which is obtained in a similar manner as the calculation performed by \citet{znajek78}, with the difference that we set the BH potential drop to the specific energy of the particles at the innermost stable orbit, whereas \citet{znajek78} makes use of the fact that the Eddington luminosity sets an upper bound on the radiation pressure (as the disk is radiatively efficient). The maximum value of the BH magnetic field corresponds to the time when the accretion rate was as high as the Eddington accretion rate. In this case, the BH spin parameter\footnote{The BH spin parameter is defined as $a_* \equiv J/J{\max}\, (= a/r_{\mathrm{g}})$, where $a=J/Mc$ is the angular momentum of the BH about the spinning axis per unit mass and per speed of light and $J_{\max} = GM^2/c$ is the maximal angular momentum of the BH. Furthermore, the BH spin parameter obeys the condition: $-1\leq a_*\leq +1$. } is limited to $a_*=0.9982$ \citep{t74}. Although this limit might be even closer to the maximal value of the spin parameter $a_*
\sim 1$, this will introduce just a small variation of the maximum value of the BH magnetic field.

\subsection{Luminosity and flux of the UHECR}
\label{lumUHECR}

In this section, we seek for the UHECR luminosity flux ($F_{\mathrm{CR}}$) specified as a function of the observed radio flux density. First, we consider the UHECR luminosity defined as
\begin{equation}
L_{\mathrm{CR}} =  \epsilon_{\mathrm{CR}} P_{\mathrm{j}} = \epsilon_{\mathrm{CR}} \gamma_ {\mathrm{j}}{\dot{M}}_{\mathrm{j}}c^2,
\label{lum}
\end{equation} 
where it is assumed that the UHECR luminosity is a fraction ($\epsilon_{\mathrm{CR}}$) of the jet power, with $P_{\mathrm{j}} = L_{\mathrm{kin}}+L_{\mathrm{magn}}+L_{\mathrm{CR}}$. If we were to adopt the point of view that the jet power is shared equally in a comoving frame between the baryonic matter, magnetic field, and cosmic rays extending to the highest energy, $\varepsilon_{\mathrm{CR}}\simeq 1/3$. In the jet-disk model of \citet{fb95}, the energy equipartition in the comoving frame appears to be a good approximation. It would also suggest that AGN driven by the BH rotational energy, and suffering from a low mass accretion rate, may attain a higher Lorentz factor, consistent with some observations.

Now, we can obtain the cosmic ray luminosity by dividing the expression of the jet power in Eq.~\ref{pjet} by 3. In section~\ref{set}, we write down the UHECR luminosity for the case of $p=2.4$. Given the UHECR luminosity, we can easily obtain the UHECR flux:
\begin{equation}
F_{\mathrm{CR}}=\frac{L_{\mathrm{CR}}}{4\pi D_{\mathrm{s}}^2},
\end{equation} 
where we do not include the cosmological distance as we refer to nearby radio sources with a flat-spectrum core and a redshift down to $z \sim 0.025$.

\subsection{Maximum particle energy of the UHECR}
\label{emax}

Now, we look for the maximum energy of the UHECR in the case of the spatial (geometrical) limit \citep{fb95}; i.e., the jet particle orbits must fit into the Larmor radius. Conform to \citet{ga99}, the maximum particle energy in the downstream rest frame can be written as
\begin{equation}
E_{\mathrm{max}}^{\mathrm{sp}} =\gamma_{\mathrm{s}} e  Z B_0 r ,
\label{gall}
\end{equation} 
where $\gamma_{\mathrm{s}}$ is the Lorentz factor of the shock and $Z$ is the particle mass number. Equation \ref{gall} corresponds to relativistic shocks, being larger by a factor $\gamma_{\mathrm{s}}$ than that resulting from a conventional geometrical comparison of the gyration radius \citep{hillas84}.

Using the expression for the magnetic field along the jet (Eqs. \ref{magnfield2} and \ref{field}) and the fact that $\tan\theta=r_0/z_0$, the maximum energy of the UHECR particles (in the observer frame) becomes:
\begin{equation}
E_{\mathrm{max}}^{\mathrm{sp}} = e Z B_{\mathrm{H}}^{\mathrm{max}} r_0 \left( \frac{\gamma_{\mathrm{s}}}{\gamma_{\mathrm{j}}}\right) \left(\frac{B_{\mathrm{H}}}{B_{\mathrm{H}}^{\mathrm{max}}} \right).
\end{equation} 
For protons,
\begin{eqnarray}
E_{\mathrm{max}}^{\mathrm{sp}} \simeq 5\times10^{20}   \left(\frac{B_{\mathrm{H}}}{B_{\mathrm{H}}^{\mathrm{max}}} \right) \left(\frac{r_0}{2r_\mathrm{g}} \right) \left( \frac{M}{10^9 M_{\odot}}\right)^{1/2} \; (\textrm{eV}),
\label{spatial}
\end{eqnarray} 
where $\gamma_{\mathrm{s}} \simeq \gamma_ {\mathrm{j}}$ was used.

Next, we look for the maximum energy of the UHECR in the case of the synchrotron loss limit \citep{bs}. Setting synchrotron losses equal to diffusive shock acceleration gains, \citet{bs} showed that a ubiquitous cutoff in the non-thermal emission spectra of AGN can be explained. This requires that the protons to be accelerated near $10^{21}$ eV. The frequency cutoff ($\nu_*$) might be produced at about $(3 - 6) \times10^{3} \,r_{\mathrm{g}}$. Rewriting the expression for the maximal proton energy derived by \citet{bs},
\begin{equation}
E_{\mathrm{max}}^{\mathrm{loss}} \simeq 1.4 \times 10^{20} \left( \frac{\nu_*}{3\times 10^{14}\mathrm{Hz}}\right)^{1/2} B^{-1/2}\; (\textrm{eV}) ,
\end{equation} 
and using the expression for the magnetic field along the jet (Eqs. \ref{magnfield2} and \ref{field}), the maximal proton energy in the loss limit reads:
\begin{equation}
\begin{split}
E_{\mathrm{max}}^{\mathrm{loss}}   \simeq  \: & 4.2\times 10^{18} \left( \frac{\nu_*}{3\times 10^{14}\mathrm{Hz}}\right)^{1/2}\left( \frac{\gamma_ {\mathrm{j}}}{5}\right) ^{1/2} \\ 
&\left(\frac{B_{\mathrm{H}}}{B_{\mathrm{H}}^{\mathrm{max}}} \right)^{-1/2}\left( \frac{M}{10^9 M_{\odot}}\right)^{1/2}\left( \frac{z}{z_0}\right)^{1/2}  \; (\textrm{eV}).
\end{split}
\label{loss}
\end{equation} 

\subsection{Model set of equations}
\label{set}

Taking $r = 2\,r_{\mathrm{g}}$ and $B_{\mathrm{H}} \simeq B_{\mathrm{H}}^{\mathrm{max}}$, the equations for the maximum particle energy in the spatial (Eq. \ref{spatial}) and loss (Eq. \ref{loss}) limits, as well as for the UHECR luminosity (Eq. \ref{lum} with \ref{pjet}) for {\it$p=2.4$}, become:
\begin{equation}
\begin{split}
E_{\mathrm{max}}^{\mathrm{sp}}   \simeq  \: & 5\times10^{20} \,Z\,  \left( \frac{M}{10^9 M_{\odot}}\right)^{1/2} \; (\textrm{eV}) ,\\
E_{\mathrm{max}}^{\mathrm{loss}}  \simeq  \: & 4.2\times 10^{18} \left( \frac{\nu_*}{3\times 10^{14}\mathrm{Hz}}\right)^{1/2}\left( \frac{\gamma_ {\mathrm{j}}}{5}\right) ^{1/2} \\
& \left( \frac{M}{10^9 M_{\odot}}\right)^{1/2}  \left( \frac{z}{z_0}\right)^{1/2} \; (\textrm{eV}) , \\
L_{\mathrm{CR}}  \simeq  \: & 1.2\times10^{39} \beta_{\mathrm{j}}  \left(1-\beta_{\mathrm{j}}  \cos\varphi \right)^{3.45}
\left( \frac{\gamma_ {\mathrm{j}}}{5}\right) ^{7.01}\left( \frac{\tan\theta}{0.05}\right)^{1.28}\\ 
&\left(\frac{F_ {\mathrm{obs}}}{\mathrm{mJy}} \right)^{1.28}\left( \frac{D_{\mathrm{s}}}{\mathrm{Mpc}}\right)^{2.56} \left(\frac{M}{10^9M_{\odot}} \right)^{-0.78} \; (\textrm{erg}\,  \textrm{s}^{-1}).
\end{split}
\label{toate}
\end{equation}
For the expression of $E_{\mathrm{max}}^{\mathrm{loss}}$ in Eq. (\ref{toate}), we can choose $(z/z_0) \sim 10^3 $, based on the results obtained by \citet{pbb09}, which show that a first large steady shock can be produced at about $z \sim 3\times10^3 \, r_{\mathrm{g}}$ [following the work by \citet{markoff01}]. Nevertheless, for the calculation of the UHECR energy spectrum, we use the expression of $E_{\mathrm{max}}^{\mathrm{sp}}$, which provides higher values for the particle maximum energy. We note that if $(z/z_0) > 10^3 $, we should make use of $E_{\mathrm{max}}^{\mathrm{loss}}$.

\section{Predictions for nearby galaxies as ultra-high-energy cosmic ray sources}
\label{sources}

In order to formulate predictions for the sources of UHECRs, we use a two-step analysis, as follows:
\begin{enumerate}
 \item we elaborate a physical model of the UHECR sources (previous section) and then
 \item we consider the deflection of the trajectories of the UHECRs by the intergalactic magnetic fields and calculate the distribution of the arriving directions and the energy spectrum of the UHECRs using both the SAM code and the CRPropa code (see below).
\end{enumerate}

We apply the model proposed in this paper to a complete sample of extended, steep spectrum radio sources\footnote{For radio galaxies, a flat-spectrum core source has a spectral index which is typically $\alpha\leqslant 0.5$ at the very center and $\alpha\simeq 0.5-0.7$ at a larger radius (the compact radio core can be extremely weak compared with the very bright and luminous radio lobes, suggesting that the radio core might suffer absorption). The latter spectral index corresponds to a power-law index of the accelerated particle of $p\simeq 2.0-2.4$. The extended emission, which includes the emission from the radio lobes, can have a steep spectrum.} \citep{biermann08,laur11a}, at redshift $z < 0.025$ (about 100 Mpc), with a total radio flux density larger than 0.5 Jy. The numbers for the estimated flux and maximal energy exclude the GZK effect, but includes the distance effect. The selection criteria used by the authors are presented in more detail in their papers. Table~\ref{29table} lists the predictions 
for 
the maximum energy and flux of the UHECRs. To calculate the errors of these quantities, we use the method of propagation of uncertainty. We emphasize that there could be a common scaling limit, such as a condition that the Larmor radius has to fit three times or five times into the jet. The scaling limit is not critical to our predictions as long as we refer the quantities to, say, those of M 87; therefore, the jet parameters ($\gamma_{\mathrm{j}}$, $\varphi$, and $\theta$) for all the sources in Table~\ref{29table} are assumed to be the same as for M 87. In the simulations described below, we use columns 5 and 7 of Table~\ref{29table}.

\begin{figure}\centering
\epsfig{file=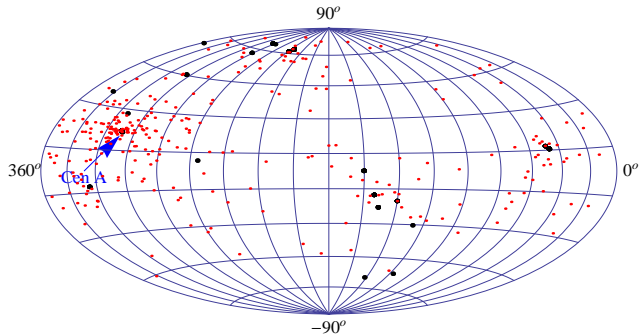,height=4.5cm}
\caption{Aitoff projection in galactic coordinate of the scattered 300 events (red dots) coming from a selected population of AGN sources (black dots). The maximal energy used in this case is taken from Table~\ref{29table}, column 5.}
\label{fig:skyplot}
\end{figure}

For the SAM code, we start with a sufficient initial number of particles at the level of the detected UHECR flux in the selected energy interval (above $2\times 10^{19}$ eV) and then distribute these particles to each source in the list (Table~\ref{29table}) taking into account the calculated flux of the UHECRs at the source. This is performed by using the Monte-Carlo method in which we consider the calculated UHECR fluxes as weights to randomly distribute the particles to each source in the list. This is equivalent to drawing numbers from a distribution according to a list of weights which is attached to the distribution. In this way, the most active source produces more UHECRs and, therefore, we can relate such source with more particles from the total number of the injected particles (that propagate through intergalactic and Galactic magnetic fields). Next, to associate a value of the energy to each UHECR, we randomly generate a range for the energy of the particles that come from each source using three 
power laws: $E^{-2.2}_{\rm inj},\, E^{-2.3}_{\rm inj}$, and $E^{-2.4}_{\rm inj}$, where we consider the estimated maximal energy of the particles as an upper limit of the energy range. Here, we assume the initial distribution of the UHECRs at the source to be constituted of 90\% protons and 10\% iron nuclei, as considered by \citet{2008JCAP...10..033A}. Future upgrade of the SAM code will include the interactions of the UHECR particles with different backgrounds (e.g., CMB, IR, Optical, and UV radiation).

To obtain the energy spectrum and the distribution of the arrival directions at the Earth of the UHECRs, we estimate the deflections in the intergalactic magnetic fields using the method applied in the numerical simulations performed by \citep{2008ApJ...682...29D}. This method provides a distribution of deflection angles of protons in the intergalactic magnetic fields corresponding to the following intervals of the energy of the particles: (i) from $10$ EeV to $30$ EeV, (ii) from $30$ EeV to $60$ EeV, and (iii) beyond $60$ EeV. When taking into account the deflection of UHECR trajectories by the magnetic fields, we have to consider separately the effects produced by the intergalactic magnetic fields \citep{1998AA...335...19R,2005JCAP...01..009D,2008Sci...320..909R} from those produced by the magnetic field of our Galaxy \citep{1991AA...245...79B,2008ApJ...674..258E,2010ApJ...711...13E}. For the SAM code, we use the scattering laws from the Fig.7 of \citet{
2008ApJ...682...29D} with the following fitting formulas:
\begin{equation*}
 N_{\theta}=\begin{cases}
 \frac{180^{o}-\theta}{(\theta_{0}^{2}+\theta^{2})^{1/2}} \; {\rm and} \; \theta_{0}=160^{o},& {\rm from} \, 10 \, {\rm EeV} \, {\rm to} \, 30 \, {\rm EeV},\\
 \frac{180^{o}-\theta}{(\theta_{0}^{2}+\theta^{2})^{1/2}} \; {\rm and} \; \theta_{0}=50^{o},& {\rm from} \, 30 \, {\rm EeV to} \, 60  \, {\rm EeV},\\
 \frac{180^{o}-\theta}{(\theta_{0}^{2}+\theta^{2})^{3}} \; {\rm and} \; \theta_{0}=40^{o},& {\rm beyond} \, 60 \, {\rm EeV},
\end{cases}
\end{equation*}
where $\theta$ denotes the deflection angle and $N_{\theta}$ represents the fraction of events per deflection angle bin, which can be easily translated into a probability to have a certain number of events with a specific deviation angle. Moreover, the core of the distribution of angle deflection is chosen such that to simulate deflections in the Galactic disk, while the rest of the distribution of the deflection angle reflects either the scattering in the cosmological magnetic fields or the scattering in a Galactic magnetic wind which extends into the halo of our Galaxy.

Again, using a Monte-Carlo weighted choice for each energy range, we associate to each UHECR a deflection angle from the direction of the source on the Earth sky, and by uniformly random selection we chose a position of the UHECR event in a ring around the source. This leads us to obtaining a sky map of the distribution of the arrival directions of the UHECRs (Fig.~\ref{fig:skyplot}). We can easily notice that almost all particles come from Cen A, which is the most active source in the list in terms of the UHECR flux. Near the position of this source, a clustering of events is observed, a picture that resembles the one obtained by the Pierre Auger Observatory \citep{auger07}. Such clear cluster of arrival directions of UHECR events around the Cen A point to their source.

To consistently check the results of the SAM code, we compare them with those obtained from simulations that we perform with the CRPropa code \citep{2013APh....42...41K}. The CRPropa code is freely available and designed to study the propagation of UHECRs in the intergalactic space by including the effects of background interactions and magnetic fields. For the simulations run with the CRPropa code, we use the same initial conditions as for those performed with the SAM code; that is, the same list of sources, initial composition, and initial flux of UHECRs. However, there are two main differences between the two simulation codes that concern (i) the interaction with the backgrounds (which is present in the CRPropa code, but not in the SAM one) and (ii) the models for the intervening magnetic fields between the sources of UHECRs and the Earth's atmosphere (for the SAM code, we consider both the extragalactic and the Galactic magnetic fields, as described above, whereas for the CRPropa code, we use the 
provided standard intergalactic magnetic field. In the present public version of the CRPropa code, the Galactic magnetic field is not included).

\begin{figure}\centering
\epsfig{file=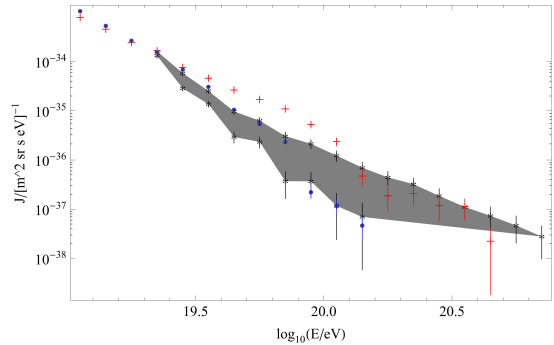,height=4.5cm}
\caption{Simulated energy spectrum produced by the LLAGN sources in Table~\ref{29table} (gray area); it represents a collection of 1000 Monte-Carlo simulations perform with the SAM code. The border points represents the minimum and maximum value per energy interval. The energy spectrum that results from the CRPropa simulations is shown with red crosses. The injected energy spectrum at the sources used in both codes (SAM and CRPropa) follows a $E_{\rm inj}^{-2.4}$ power-law. The UHECR energy spectrum of the Pierre Auger Observatory \citep{2011arXiv1107.4809T} is represented in blue dots.}
\label{fig:energyspectrum}
\end{figure}

Figure~\ref{fig:energyspectrum} shows the energy spectrum resulted from the simulations performed with the SAM code (the grey area), which is compared with that obtained with the CRPropa code (red crosses) and with the observed by the Pierre Auger Collaboration (blue dots). Here, we show only the plot corresponding to the injected energy spectrum of particles of $E_{\rm inj}^{-2.4}$, which provides the best approximation to the observed spectrum. The other two slopes ($p = 2.2$ and $p = 2.3$) generate  much higher energy spectra for UHECRs than $p=2.4$. In order to estimate the simulation errors, a set of 1000 simulations are performed and the global error is computed according to \ref{MCerrors}. Here, we use from the all sky distribution of arriving particles only the ones which fall on the general field of view of the Pierre Auger detector. We note that the experiment makes a certain cut-off in the events that it detects which depends on many parameters like the arrival angle of the particles with 
respect of the Pierre Auger detector or other factors which are not considered here (for a discussion see \citet{2011arXiv1107.4809T}). These factors could explain the large errors that appear at the highest energies, as well as the offset of the end points. However, a slight overestimate of the maximal energy of UHECRs at the source might also account for the large errors at the highest energies in the spectrum. Within errors, the simulation results obtained with the SAM code are in good agreement with the observed energy spectrum. The CRPropa simulation provides an energy spectrum which, for the energy range $E\in[10^{19.5},10^{20}]$ eV, is slightly above both the SAM energy spectrum and the observed one. We correlate this discrepancy with the fact the CRPropa code does not include deflection of particles in the Galactic magnetic field.

\begin{figure}\centering
\epsfig{file=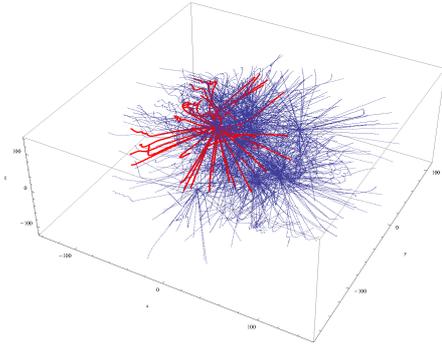,height=4.5cm}
\caption{3D representation of UHECR particles which are generated from the LLAGN listed in Table~\ref{29table}. The red curves represent the trajectories of the UHECR particles generated from Cen A. The simulation box size is $100\times100\times100$ Mpc$^3$, where our Galaxy is located at the center of the box.  Here, we use the CRPropa simulation code.}
\label{fig:3D}
\end{figure}

Using a special developed representation software in Mathematica, we also include here a 3D representation of the trajectories of UHECRs (Fig.~\ref{fig:3D}), generated from the 29 LLAGN in Table~\ref{29table}. We obtain this representation from simulations with the CRPropa code where we use the same initial setup as for the SAM code. We marked in red the trajectories of the particles that are generated from Cen A. The simulation box size is $100\times100\times100$ Mpc$^3$, where our Galaxy is located at the center of the box.

\section{Summary and conclusions} 
\label{sec:summary}

In this paper, we address the possibility that nearby LLAGN can be sites of production of UHECRs. The analysis that we perform in the work presented here is two-fold:
\begin{enumerate}
 \item First, we elaborate a physical model of the source of UHECRs, where the particles are accelerated in LLAGN jets, for which the acceleration regions can be sustained by the  
       magnetic energy extraction from the spinning BH at the heart of the AGN.  We relate the observed radio flux density to the luminosity and flux of the UHECRs and calculate the maximum particle energy in both spatial and loss limits. Next, we calculate the energy and flux of UHECR that can be generated from a complete sample of 29 LLAGN (Table~\ref{29table}). The results of these calculations are then used as initial conditions for the simulations performed with both the SAM code and the CRPropa one.
 \item Second, using the SAM code we estimate the deflection of the trajectories of the UHECRs (90\% protons and 10\% iron nuclei) by the intergalactic and Galactic magnetic fields, using appropriate distributions of the angle deflections (Section \ref{sources}) and calculate the distribution of the arriving directions and the energy spectrum of the UHECR particles. We also compare the energy spectrum obtained from the SAM simulations with that resulted from the simulations with the CRPropa code and from the measurements performed by the Pierre Auger Observatory. 
 
 \end{enumerate}

From the simulations performed with the SAM code and CRPropa, the best approximation to the observed spectrum measured by the Pierre Auger Observatory is obtained for the injected energy spectrum of particles corresponding to $E_{\rm inj}^{-2.4}$. Although the two simulated spectra are in good agreement with the observed spectrum at the highest energies, the spectrum obtained with the CRPropa code overestimates the one measured by the Pierre Auger Observatory in the energy range $E^{-19.5}-E^{-20}$ eV. The different behavior of the simulated energy spectra care be attributed to the differences between the two codes. The main difference between the codes regards the deflection of the particles in the Galactic magnetic field, which is not included in the current public version of the CRPropa code, whereas such deflection is already implemented in the SAM code. More importantly, to calculate the energy spectrum, we also include the weighting of the UHECR flux per each source; i.e., we attach to each UHECR 
source a weigh in flux based on the value of its UHECR flux relative to a canonical value of the flux, which we choose it to be the flux corresponding to M 87. (The use of the relative flux to a canonical value can eliminate the systematic errors.) This weighting of the UHECR flux is not taken into account in the CRPropa code, where all sources can have the same UHECR flux. Therefore, as in the CRPropa code there is not possible to run a single simulation with different number of particles per each source, we have to run separate simulations per each UHECR source (or sublist of sources) from the sample in Table~\ref{29table}. Thus, we run (i) one simulation for Cen A using $60\%$ of the total injected particle number ($N_{\rm inj}$), (ii) one simulation for M 87 and the sources with an UHECR flux at the same level of that of M 87 using $32\%$ of $N_{\rm inj}$, and (iii) one simulation for the rest of the sources in Table~\ref{29table} using the remaining $8\%$ of $N_{\rm inj}$. Next, we merge these three 
simulations together to obtain the energy spectrum of UHECRs. The SAM code has the weighting of the UHECR flux already implemented, so that we just use the UHECR flux provided by the model to give the initial number of particle per source. 

Although we presented a simple model for relativistic jets in LLAGN as sites of production of UHECRs, without including complex phenomena (e.g., magnetic reconnection and instabilities in the jet plasma) that can change the spectrum of the emitted radiation from the jets, the uncertainty on several parameters of the model are still present. For example, we do not know exactly at which distance from the BH a strong shock can be produced. We might overcome this problem by studying each source individually and using observations of high resolution, which may not be available. However, in this paper we have to make assumptions based on the performed measurements. We also do not have observational data for the jet parameters (the Lorentz factor, the angle to the line of site, and the semi-opening angle) of the all sources in Table~\ref{29table}. We supposed that these jet parameters have similar values as those of M 87. (We note that the difference between the value of the UHECR flux using the observed parameters 
of the jet for Cen A and that for Cen A but using the parameters of the jet for M 87 is within one order of magnitude.) Thus, the maximum particle energy and flux of UHECRs will be scale only by the BH mass, the radio power, and the distance to the source:

\begin{equation*}
 E_{\mathrm{max}}^{\mathrm{sp}} \sim \left( \frac{M}{10^9 M_{\odot}}\right)^{1/2},
 \end{equation*} 

\begin{equation*}
 F_{\mathrm{CR}} \sim \left(\frac{F_ {\mathrm{obs}}}{\rm mJy}\right)^{\frac{p+4}{5}} \, \left(\frac{D_{\mathrm{s}}}{\rm Mpc} \right)^{\frac{2(p-1)}{5}}   \left(\frac{M}{10^9M_{\odot}} \right)^{-\frac{2p+3}{10}}.
\end{equation*}
where the errors are given by the uncertainties in the radio core flux density, the distance to the AGN, and the BH mass measurements. Therefore high resolution observations in different frequency domains are needed. Moreover, increasing the number of BHs with measured mass, the distribution function of the BH mass of \citet{laur10,laur11b} can be improved and then use to determine the masses of supermassive BHs more precisely. 

We plan to extend the analysis presented in this paper. We will implement the calculations for neutrino and gamma-ray energy spectra. We also intend to use a different sample of LLAGN, e.g., the catalog by \citet{vanVelzen2012} and to modify the chemical composition of the UHECRs. These calculations can provide a further test on the exact composition of the UHECRs as different particles behave differently when they propagates through the intergalactic and Galactic magnetic fields.

\section*{Acknowledgments}
We would like to thank Peter L. Biermann. We also thank the anonymous referees for their thorough review and highly appreciate their comments and suggestions, which significantly contributed to improving the quality of the paper. I.D. was supported through a stipend from the International Max Planck Research School (IMPRS) for Astronomy and Astrophysics at the Universities of Bonn and Cologne. She appreciates the support from MPIfR during the last phase of this work. I.D. was also supported by STAR, CONTR. 12/19.11.2012, and L.I.C. was partially supported by PN-II-RU-TE-2011-3-0184 and STAR, CONTR. 12/19.11.2012. We make use of the NASA/IPAC Extragalactic Database (ned.ipac.caltech.edu) and the CRPropa simulation code (https://crpropa.desy.de/Main\_Page).

\appendix
\section{Electron and proton number densities}
First, we consider the case that the jet is composed mainly of electrons, positrons, and protons. Then, we chose the case where the mass composition of the UHECRs consists of 90\% protons and 10\% iron nuclei instead of 100\% protrons. We denote by  $f_{\mathrm{ep}} \equiv n_{\mathrm{e}}/n_{\mathrm{p}}$ the ratio of the electron to proton number densities, where the number densities are measured in a frame comoving with the jet plasma. Unless otherwise noted, $n_{\mathrm{e}}$ should be assumed to include the positron number density as well. It is straightforward to generalize to a mixed chemical composition, including many heavy nuclei. Furthermore, both electrons and protons can have thermal and non-thermal populations before being accelerated at the shock. There may also be a substantial number of positrons from pion production and decay processes (also called secondaries).

Now, we look for the expression of the proton and electron number densities injected into the accelerating region. First, we consider the mass flow rate into the jets, which in the comoving frame is given by
\begin{equation}
\dot{M}_{{\mathrm{j,co}}} =\frac{d}{dt}\left( \rho_{\mathrm{j}} V_{\mathrm{j}} \right) =\frac{d}{dt} \left[ n  m  z (S)_{z=0} \right] = n  m  v_{\mathrm{j}}(S)_{z=0} ,
\label{mjet}
\end{equation} 
where $\rho_{\mathrm{j}} $ is the rest-mass density of the jet, $V_{\mathrm{j}} $ is the comoving volume of the jet, $(S)_{z=0}$ is the launching area of the jet, $z$ is the length of the cylinder along which the jet propagates before expanding freely in a conical geometry, and  $v_{\mathrm{j}} = \beta_ {\mathrm{j}} c $ is the bulk velocity of the jet.

The surface area between two equatorial surfaces of a Kerr BH can be calculated as 
\begin{equation}
(dS)_{z=0}=\left( \frac{A}{\Delta}\right) ^{1/2}2\pi dr ,
\end{equation} 
where the Kerr metric functions are:
\begin{equation}
\Delta =r^2-2r_\mathrm{g}r+a^2 \ \mathrm{and}\ A=r^4+r^2a^2+2r_\mathrm{g}ra^2 ,
\label{coeff}
\end{equation}
where $r$ is the coordinate radius. Next, we use normalizations to the gravitational radius, so that $r_*=r/r_{\mathrm{g}}$ is the dimensionless radius. The surface area is then:
\begin{equation}
(S)_{z=0} = 2\pi r_{\mathrm{g}}^2 \int_{r_{\mathrm{ms}_*}}^{r_{\mathrm{sl}_*}}\limits r_* \sqrt{\frac{1 + r_*^{-2}a_*^2 + 2r_*^{-3}a_*^2}{1 - r_*^{-1} + r_*^{-2}a_*^2}}dr_* \equiv  2\pi r_{\mathrm{g}}^2 k_0 ,
\label{aria}
\end{equation} 
where the factor $k_0$ increases from $\sim$ 2 to $\sim$ 80 as the BH spin parameter increases from 0.95 to $\sim$ 1. For the first equality, we use the fact that the inner disk, from where the jet is launched, has its inner and outer radii at the innermost stable orbit\footnote{Once the accretion flow reaches the innermost stable orbit, it drops out of the disk and falls directly into the BH. The expression for the radius of the innermost stable orbit $r_{\mathrm{ms}}$ is given by eq. (2.21) in \citet{bardeen70}.} $r_{\mathrm{ms}}$ and stationary limit $r_{\mathrm{sl}}= 2 r_{\mathrm{g}} \equiv r_0$, respectively. 

The comoving density of the jet can be expressed in terms of the ratio of the electron to proton number densities:
\begin{equation}
n  m=n_{\mathrm{p}} m_{\mathrm{p}} + n_{\mathrm{e}} m_{\mathrm{e}} = n_{\mathrm{p}} m_{\mathrm{p}} \left( 1+ f_{\mathrm{ep}} \frac{m_{\mathrm{e}}}{m_{\mathrm{p}}}\right) \equiv n_{\mathrm{p}} m_{\mathrm{p}}f_0 .
\label{density}
\end{equation} 
For protons dominating over the electrons, $f_{\mathrm{ep}} < 2\times10^{3}$, where electrons and positrons can partially occur as secondaries. This fraction of CR is $\sim 10^{-2}$ from data of CRs at 1 GeV. One can get a ratio of unity assuming that the spectra go down to rest mass, which is implausible [see, e.g., \citet{protheroe96} and references therein].

Substituting Eqs. (\ref{aria}) and (\ref{density}) for (\ref{mjet}), we obtain the mass flow rate into the jet in the observer frame by including $\gamma_{\mathrm{j}}$:
\begin{equation}
\dot{M}_{{\mathrm{j}}} =\gamma_{\mathrm{j}} \beta_ {\mathrm{j}} c   n_{\mathrm{p}} m_{\mathrm{p}} f_0   2 \pi r_{\mathrm{g}}^2 k_0.,
\label{eq:mdot}
\end{equation} 
This expression provides the proton number density, which we use to derive the electron number density:
\begin{equation}
n_{\mathrm{e}}=f_{\mathrm{ep}} \frac{\dot{M}_{{\mathrm{j}}}}{\gamma_{\mathrm{j}} \beta_ {\mathrm{j}} c   m_{\mathrm{p}} f_0   2 \pi r_{\mathrm{g}}^2 k_0 }.
\label{edensity}
\end{equation} 
Now, if we consider that the UHECR particles are composed of 90\% protons and 10\% iron nuclei, the denominator of Eq.~\ref{edensity} is modified through a multiplying factor of $\sim 7$, as we replace $n_{\mathrm{p}} m_{\mathrm{p}}$ with $n_{\mathrm{p}} m_{\mathrm{p}} + n_{\mathrm{Fe}} m_{\mathrm{Fe}}$. We shall use this result later for evaluating the self-absorbed synchrotron emission of the jets (Section \ref{secSAS}).

\section{Self-absorbed synchrotron emission of the jets}
\label{secSAS}

The spectra from compact radio sources can be explained by self-absorbed synchrotron emission of the jets produced by electrons with a power-law energy distribution. In this section, we rewrite the quantities which describe the self-absorbed synchrotron emission, derived in \citet{ryb+ligh}, and express them under the considerations of the model presented here. Using Eqs. \ref{norm2} and \ref{magnfield2}, the absorption coefficient [eq. 6.26 in \citet{ryb+ligh}] becomes:
\begin{equation}
\alpha_{\nu} = K_1 C'_0   \left( \frac{z}{z_0}\right)^{-\frac{p+6}{2}}    B_0^{\frac{p+2}{2}}   \nu^{-\frac{p+4}{2}} ,
\label{absorptionFin}
\end{equation} 
where
\begin{equation}
\begin{split}
K_1 \simeq \: & 8.4\times10^{-3} (1.25\times10^{19})^{\frac{p}{2}}  \left( 8.2\times 10^{-7}\right)^{p-1} \frac{\sqrt{\pi}}{2} \\
&\Gamma\left( \frac{3p+2}{12}\right)  \Gamma\left(  \frac{3p+22}{12}\right)   \Gamma\left( \frac{p+6}{4}\right)  \Gamma^{-1}\left(  \frac{p+8}{4}\right).
\end{split}
\end{equation} 

To calculate the observed distance along the jet where the jet becomes self-absorbed, we first determine the optical depth $\tau_{\nu}$ of the jet material. The averaged path of a photon through the jet has the length $r(z)$, which is a reasonable approximation for a jet observed at large inclination angle \cite[e.g.,][]{kaiser}. We introduce a factor $l_{0}$ in the expression of the path length to account for a small inclination angle. Thus, we can write the optical depth as
\begin{equation}
\tau_{\nu}=\alpha_{\nu} r(z)l_{0} .
\end{equation} 
For conical jets, the intrinsic half-opening angle is given by $\tan\theta=r/z \cong r_0/z_0$. With the absorption coefficient specified through Eq. (\ref{absorptionFin}), the optical depth can be written as
\begin{equation}
\tau_{\nu} = K_1  C'_0  r_0  l_0  \left( \frac{z}{z_0}\right) ^{-\frac{p+4}{2}} B_0^{\frac{p+2}{2}} \nu^{-\frac{p+4}{2}} ,
\label{tau}
\end{equation} 

One can define the distance along the jet where the jet becomes self-absorbed $z_{ssa}$ as the distance $z$ for which $\tau_{\nu}=1$. Using Eq. (\ref{tau}), one obtains:
\begin{equation}
z_{ssa} = \left(  K_1  C'_0  l_0\right) ^{\frac{2}{p+4}}\left(\tan\theta \right)^{-1}  r_0^{\frac{p+6}{p+4}} B_0^{\frac{p+2}{p+4} \nu^{-1}} .
\label{zet}
\end{equation} 

The total power radiated per unit volume per unit frequency by a non-thermal particle distribution equals:
\begin{equation}
\begin{split}
P_{\omega}=  \: & \frac{\sqrt{3} e^3}{2\pi  m_ {\mathrm{e}}c^2}\frac{C' B \sin\alpha_0}{p+1} \left( \frac{m_{\mathrm{e}} c \omega}{3 e B \sin\alpha_0}\right)^{-\frac{p-1}{2}}\\
&\Gamma\left(\frac{p}{4}+\frac{19}{12} \right)\Gamma\left(\frac{p}{4}-\frac{1}{12} \right) ,
\end{split}
\end{equation} 
where $\omega=2\pi\nu$ [Eq. 6.36 in \citet{ryb+ligh}]. Using Eqs. (\ref{magnfield2}) and  (\ref{norm2}), as well as the method to calculate the averaged pitch angle employed in \citet{longair}, the expression of the total power becomes:
\begin{equation}
P_{\nu} = 2\pi P_{\omega}= K_2   C'_0   \left( \frac{z}{z_0} \right)^{-\frac{p+5}{2}} B_0^{\frac{p+1}{2}} \nu^{-\frac{p-1}{2}} ,
\label{Ptot}
\end{equation}  
where
\begin{equation}
\begin{split}
K_2 \simeq \: & 3.7\times10^{-23}  \left( 1.2\times10^{-7}\right) ^{-\frac{p-1}{2}}(p+1)^{-1}\frac{\sqrt{\pi}}{2} \\ 
& \Gamma\left(\frac{p}{4}+\frac{19}{12} \right)\Gamma\left(\frac{p}{4}-\frac{1}{12} \right)\Gamma\left(\frac{p+5}{4}\right)\Gamma^{-1}\left(\frac{p+7}{4}\right) .
\end{split}
\end{equation}
Next, we derive $(z/z_0)$ from Eq. \ref{tau} when $\tau_{\nu}=1$. With this, the expression for the total power takes the form:
\begin{equation}
P'_{\nu} = K_2 \left(K_1  r_0  l_0\right)^{-\frac{p+5}{p+4}} (C'_0)^{-\frac{1}{p+4}} B_0^{-\frac{p+3}{p+4}} \nu^3,
\end{equation} 
and the emission coefficient is simply $j_{\nu} = P_{\nu}/4\pi$. 

At low frequencies, the emitting region is opaque to synchrotron radiation and the observed intensity of radiation is proportional to the source function, while at high frequencies, the region is transparent and the observed intensity is proportional to the emission coefficient. This transition corresponds to an optical depth $\tau_{\nu} = 1$.  Using Eqs. (\ref{absorptionFin}) and (\ref{Ptot}), the source function ($S_{\nu} = P_{\nu}/(4\pi\alpha_{\nu})$) when $\tau_{\nu}=1$ becomes:
\begin{equation}
S'_{\nu}=K_3  \left( C'_0  r_0  l_0\right)^{\frac{-1}{p+4}}  B_0^{-\frac{1}{p+4}}\nu^2 \;(\textrm{erg s}^{-1}\textrm{cm}^{-2}\textrm{Hz}^{-1}) ,
\label{snu}
\end{equation} 
where $K_3 = K_1^{-\frac{p+3}{p+4}}K_2$.

To obtain the emission spectrum, one needs to solve the equation for the radiative transfer through a homogeneous medium. Because the angular sizes of the jets are small, instead of the specific intensity of the radiation, one usually measures the flux density $F_{\nu}$ (energy per unit time, per unit frequency interval, that passes through a surface of unit area). Thus,
\begin{equation}
dF_{\nu} = I_{\nu}d\Omega = S_{\nu}\left[ 1-\exp(-\tau_{\nu})\right] d\Omega,
\label{df}
\end{equation} 
where $d\Omega =2\pi r dz/D_{\mathrm{s}}^2$, with $D_{\mathrm{s}}$ the distance from the observer to the jet source and $r= z \tan\theta$. If we integrate Eq. \ref{df} from $z_0$ to $z$, we obtain the flux density of the synchrotron emission in the case of $\tau_{\nu}=1$ as
\begin{equation}
F'_{\nu} = S'_{\nu}[1-\exp(-1)]  \pi (\tan\theta) D_{\mathrm{s}}^{-2} z^2\left[1-\left(\frac{z_0}{z} \right)^{2} \right] ,
\end{equation} 
where the second term in the last squared bracket can be neglected with respect to the first term for $z \gg z_0$ (where the jet emission becomes self-absorbed). Using Eqs. (\ref{zet}) and (\ref{snu}), the flux density is then: 
\begin{equation}
F' = K_4  (C'_0  l_0)^{\frac{5}{p+4}}   r_0^{\frac{2p+13}{p+4}} B_0^{\frac{2p+3}{p+4}} D_{\mathrm{s}}^{-2}  (\tan\theta)^{-1},
\label{flux}
\end{equation} 
where $K_4 \simeq 0.16   K_1^{-\frac{p-1}{p+4}}K_2$. The radio flux density in Eq. (\ref{flux}) does not depend on the emitted frequency of the radiation since we already adopted the case of flat-spectrum core sources when $\tau_{\nu}=1$. 

For a power-law synchrotron spectrum of the form $F_{\mathrm{obs}}\sim \nu_{\mathrm{obs}}^{-\alpha}$ from a continuous jet, the observed flux density is related to the intrinsic flux density as
\begin{equation}
F_{\mathrm{obs}}=\mathcal{D}_{\mathrm{j}}^{2+\alpha} F' ,
\label{cblobs}
\end{equation} 
where $\mathcal{D}_{\mathrm{j}}=\gamma_{\mathrm{j}}^{-1}(1-\beta_{\mathrm{j}}\cos\varphi)^{-1}$ is the Doppler factor of the jet and $\varphi$ is the inclination angle of the jet axis with respect to the line of sight .

\section{Relation between the jet power and the observed radio flux density for a flat-spectrum core source}
\label{jetpower}

In the previous section, we established the expression for the radio flux density from sources with a flat-spectrum core (Eq. \ref{flux}). This quantity reflects the radiative property of the jet, as the radiated energy is replaced by dissipation of the jet power \citep[e.g.,][]{bk79}. In this section, we seek the relation between the jet power and the observed radio flux density. First, we consider the jet power in the observer frame defined as
\begin{equation}
P_{\mathrm{j}} =  \gamma_ {\mathrm{j}}{\dot{M}}_{\mathrm{j}}c^2,
\label{jet}
\end{equation} 
which follows, e.g., from \citet{fb95} [see also \citet{vila10}], and for which we need to evaluate ${\dot{M}}_{\mathrm{j}}$ using Eq. \ref{edensity}. An upper limit for the electron density is specified by $n_{\mathrm{e}} \leqslant C'_0 $. So, we can substitute Eq. (\ref{edensity}) for the expression of the observed radio flux density (Eq.~\ref{cblobs}) and find the mass flow rate into the jet ${\dot{M}}_{\mathrm{j}}$. The strength of the magnetic field $B_0$ follows from Eqs. (\ref{field}) and (\ref{maxB}). This procedure yields the power of the jet: 
\begin{equation}
\begin{split}
P_{\mathrm{j}} =  & K_5 f   \beta_ {\mathrm{j}}(1-\beta_{\mathrm{j}}\cos\varphi)^{-h} \left( \frac{\gamma_ {\mathrm{j}}}{5}\right) ^{\frac{2p+13}{5}+h} \left( \frac{\tan\theta}{0.05}\right)^{\frac{p+4}{5}} \left( \frac{r_0}{2r_{\mathrm{g}}}\right) ^{-\frac{2p+13}{5}}  \\
&\left(\frac{B_{\mathrm{H}}}{B_{\mathrm{H}}^{\mathrm{max}}} \right)^{-\frac{2p+3}{5}}  \left(\frac{F_ {\mathrm{obs}}}{\rm mJy}\right)^{\frac{p+4}{5}} \, \left(\frac{D_{\mathrm{s}}}{\rm Mpc} \right)^{\frac{2(p+4)}{5}}   \left(\frac{M}{10^9M_{\odot}} \right)^{-\frac{2p+3}{10}}    \textrm{erg} \textrm{s}^{-1} ,
\label{pjet}
\end{split}
\end{equation}  
where 
\begin{equation}
\begin{split}
K_5 \simeq  \: & \frac{\pi}{2} m_{\mathrm{p}} c^3 K_4^{-\frac{p+4}{5}} (5)^{\frac{2p+13}{5}+h}(0.05)^{\frac{p+4}{5}}\\
& (2.96 \times 10^{14})^{-\frac{2p+3}{5}} (0.56 \times 10^{4})^{-\frac{2p+3}{5}} \\
& (3\times10^{24})^{\frac{p+4}{5}}(10^{-26})^{\frac{2(p+4)}{5}},
\end{split}
\end{equation} 
where $h = [(p+3)(p+4)]/10$ for a continuous jet emission (Eq. \ref{cblobs}) and $ f = f_0  k_0  (l_0  f_{\mathrm{ep}})^{-1}$.  We use a normalization value for the Lorentz factor of the jet, say 5, although this factor can range from $\sim 2$ to $\sim 100$, as observational data suggest. We adopt $f_{\mathrm{ep}} \sim 10^{-2} $ (and then $f_0 \simeq 1$), which means that there is, in average, one hundred electrons/positrons for at least one proton (or one heavy nucleus) in a jet that is powered by a very rapidly spinning BH ($a_* \geqslant 0.95$) and observed at a large angle ($\geqslant 10^{\circ}$). Flat spectrum cores are predicted for any angle to the line of sight \citep{bk79}. They are pointing close to the line of sight only if the cores dominate over the extended emission.

Using Eq. (\ref{pjet}) for the observed radio flux density of a conical jet, we obtain:
\begin{equation}
F_ {\mathrm{obs}} \sim  P_{\mathrm{j}}^{\frac{5}{p+4}}\, D_{\mathrm{s}}^{-2}\,M^{\frac{2p+3}{2(p+4)}}\, \mathcal{D}_{\mathrm{j}}^{\frac{p+3}{2}}\, \gamma_ {\mathrm{j}}^{-\frac{2p+13}{p+4}}(\tan\theta)^{-1},
\label{fpdm}
\end{equation} 
where the electron number density in the jet scales as $\sim \gamma_{\mathrm{j}} z^{-2}$, $B_{\mathrm{H}} \simeq B_{\mathrm{H}}^{\mathrm{max}}$, and $r_0 = 2\,r_{\rm g}$. Since the observed radio flux density in Eq. (\ref{fpdm}) is not dependent on the distance along the jet, the expression can be applied to microquasars as well.

\section{Error bars in Monte Carlo computations}
\label{MCerrors}

Monte Carlo computations are likely to have large errors and it is important to estimate the order of magnitude of the error in an easy way. Therefore, all Monte Carlo computations should report error estimates.

If we take $I$ as a random variable for $n$ times and then we consider the value of $A = X[I]$ by
\begin{equation}
\hat{A}_{n}=\frac{1}{n}\sum_{i=1}^{n}I_{i}.
\label{MC1}
\end{equation} 

Next, if we use the central limit theorem from the probability theory, which states that the mean of a sufficiently large number of independent random variables, each with a well-defined mean and well-defined variance, will be approximately normally distributed, then:

\begin{equation}
{M}_{n}=\hat{A}_{n}-A \approx \sigma N ,
\label{MC2}
\end{equation} 
where $\sigma$ is the standard deviation of $\hat{A}_{n}$ and $N \in (0,1)$. We calculate further that $\sigma$ is $ \sigma=\frac{1}{\sqrt{n}}\sqrt{var(I)}$, where $var(I) = X[(I-A)^{2}]$. Then:

\begin{equation}
\widehat{\sigma^{2}}=\frac{1}{n}\sum_{i=1}^{n}(I_{n}-\hat{A}_{n})^{2},
\label{MC3}
\end{equation} 

and finally,

\begin{equation}
\hat{\sigma}=\frac{1}{\sqrt{n}}\sqrt{\widehat{\sigma^{2}}}=\frac{1}{n}\sqrt{\sum_{i=1}^{n}(I_{n}-\hat{A}_{n})^{2}}.
\label{MC3}
\end{equation} 

The Monte Carlo data is usually described like $A=\hat{A_{n}} \pm \hat{\sigma}$ and graphically we represent the error using a bar of length $2\hat{\sigma}$ with the estimated value of $A$, $\hat{A_{n}}$ in the center and we evaluate the value of $A$ in the interval $[\hat{A_{n}} - \hat{\sigma},\hat{A_{n}} + \hat{\sigma}]$

\begin{table}
\centering
\caption{UHECR predictions for a complete sample of 29 extended, steep spectrum sources \citep{biermann08,laur11a}.}
\begin{threeparttable}
\begin{tabular}{l  c c c c c c c}
\hline
\hline
Source & D & M & $F_{\mathrm{core}}^{5\mathrm{GHz}}$ & $ E_{\mathrm{max}}^{\mathrm{sp}} $ & $ E_{\mathrm{max}}^{\mathrm{sp}} $/ & $ F_{\mathrm{CR}}$  \\

& (Mpc)& ($\times10^9 M_\odot $)&(mJy)& ($10^{21}$ eV)   &    $E_{\mathrm{max}}^{\mathrm{sp,M87}}$ & $F_{\mathrm{CR}}^{\mathrm{M87}}$   \\

\begin{small}(1)\end{small}&\begin{small}(2)\end{small}&\begin{small}(3)\end{small}&\begin{small}(4)\end{small}&\begin{small}(5)\end{small}&\begin{small}(6)\end{small}&\begin{small}(7)\end{small}\\[0.5ex]
\hline
   ARP 308 & $69.7\pm4.9$ &$0.10\pm0.05$ &$67.8\pm6.78$ &4.08$\pm$1.02&0.17$\pm$0.06&0.28$\pm$0.04\\
   CGCG 114-025&$67.4\pm4.7$&$0.19\pm0.09$ &$652.48\pm95.81$ &5.63$\pm$1.33&0.23$\pm$0.09&3.12$\pm$0.04\\
   ESO 137-G006 & $75.8\pm5.3$&$0.92\pm0.46$ & $1201.25\pm92.87$  &12.4$\pm$3.1&0.52$\pm$0.20&2.13$\pm$0.26\\
   IC 4296&$54.9\pm3.9$ &$1.00\pm0.50$&$214\pm21.4$& 12.92$\pm$3.23&0.54$\pm$0.21&0.18$\pm$0.02 \\
   IC 5063& $44.9\pm3.1$ &$0.20\pm0.10$& $237.56\pm18.28$& 5.78$\pm$1.44&0.24$\pm$0.09&0.65$\pm$0.08\\ 
   NGC 0193& $55.8\pm3.9$&$0.20\pm0.10$& $40\pm0.9 $& 5.78$\pm$1.44&0.24$\pm$0.96&0.07$\pm$0.004\\
   NGC 0383& $65.8\pm4.6$&$0.55\pm0.27$& $89\pm1.4 $& 9.58$\pm$2.35&0.40$\pm$0.15&0.10$\pm$0.004\\
   NGC 1128& $92.2\pm6.5$&$0.20\pm0.10$& $39\pm3.9$ & 5.78$\pm$1.44&0.24$\pm$0.09&0.09$\pm$0.01\\
   NGC 1167& $65.2\pm4.6$&$0.46\pm0.23$& $44.9\pm4.4$& 8.76$\pm$2.19&0.36$\pm$0.14&0.05$\pm$0.007\\
   NGC 1316& $22.6\pm1.6$&$0.92\pm0.46$& $26\pm2.6$& 12.4$\pm$3.1&0.52$\pm$0.20&0.008$\pm$0.0001\\
   NGC 1399& $18.2\pm1.3$&$0.30\pm0.15$& $10\pm1.0$& 7.08$\pm$1.77&0.29$\pm$0.11&0.005$\pm$0.0007\\
   NGC 2663& $32.5\pm2.3$&$0.61\pm0.30$& $160\pm16$& 10.09$\pm$2.48&0.42$\pm$0.16&0.13$\pm$0.02\\
   NGC 3801& $50.0\pm3.5$&$0.22\pm0.11$& $635\pm95.25$& 6.06$\pm$1.51&0.25$\pm$0.10&2.27$\pm$0.49\\
   NGC 3862& $93.7\pm6.6$&$0.44\pm0.22$& $1674\pm251$ & 8.57$\pm$2.14&0.35$\pm$0.14&6.52$\pm$1.41\\
   NGC 4261& $35.4\pm2.5$&$0.52^{+0.10}_{-0.11}$& $300\pm30$& 9.32$\pm$0.89&0.39$\pm$0.09&0.36$\pm$0.05\\
   NGC 4374& $19.1\pm1.4$&$1.0^{+2.0}_{-0.6}$& $168.7\pm0.1$& 12.92$\pm$1.29&0.54$\pm$0.13&0.07$\pm$0.001\\
   NGC 4486& $16.5\pm1.2$&$3.40\pm1.00$&$2875.1\pm0.1$& 23.84$\pm$3.5&1&1 \\
   NGC 4651& $15.1\pm1.1$&$0.04\pm0.02$&$15\pm1.5$& 2.58$\pm$0.64&0.10$\pm$0.04&0.03$\pm$0.005\\
   NGC 4696& $44.4\pm3.1$&$0.30\pm0.15$&$55\pm5.5$& 7.08$\pm$1.77&0.29$\pm$0.11&0.07$\pm$0.01\\
   NGC 5090& $50.4\pm3.5$&$0.74\pm0.37$&$268\pm26.8$& 11.12$\pm$2.78&0.46$\pm$0.18&0.29$\pm$0.04\\  
   NGC 5128& $3.5\pm0.35$&$0.055\pm0.03$&$6984\pm698.4$& 3.03$\pm$0.82&0.12$\pm$0.05&32.60$\pm$5.25\\   
   NGC 5532& $104.8\pm7.3$&$1.08\pm0.54$& $77\pm7.7$& 13.43$\pm$3.35&0.56$\pm$0.22&0.06$\pm$0.01\\
   NGC 5793& $50.6\pm3.7$&$0.14\pm0.07$& $239.61\pm27.28$&4.83$\pm$1.2&0.20$\pm$0.08&0.93$\pm$0.16\\
   NGC 7075& $72.7\pm5.1$&$0.25\pm0.12$& $46.29\pm3.42$& 6.46$\pm$1.55&0.27$\pm$0.10&0.08$\pm$0.01\\
   UGC 01841& $84.4\pm5.9$&$0.10\pm0.05$& $182\pm18.2$& 4.08$\pm$1.02&0.17$\pm$0.06&1.14$\pm$0.17 \\
   UGC 02783& $82.6\pm5.8$ &$0.42\pm0.21$&$285\pm28.5$& 8.37$\pm$2.09&0.35$\pm$0.13&0.65$\pm$0.09\\
   UGC 11294& $63.6\pm4.5$ &$0.29\pm0.14$& $12.23\pm1.37$& 6.96$\pm$1.68&0.29$\pm$0.11&0.01$\pm$0.002\\   
   VV 201& $66.2\pm4.6$& $0.10\pm0.05$& $88\pm8.8$& 4.08$\pm$1.02&0.17$\pm$0.06&0.39$\pm$0.05\\ 
   WEIN 45& $84.6\pm6.1 $&$0.27\pm0.13$& $486.15\pm38.04$& 6.71$\pm$1.61&0.28$\pm$0.10&1.85$\pm$0.23\\
\hline
\end{tabular}
\scriptsize
\hfill\parbox[t]{16cm}{NOTE: Col. (1) Source name; Col. (2) Distance to the source; Col. (3) BH mass; Col. (4) Core flux density at 5 GHz; Col (5) Maximum particle energy (spatial limit) for $Z=26$; Col. (6) Maximum particle energy relative to that of M 87 (spatial limit);  (7) UHECR flux relative to that of M 87. To calculate the errors for the quantities in columns (5), (6), and (7), we use the method of propagation of uncertainty. REFERENCES for Table~\ref{29table}: For Col. (2): The NASA/IPAC Extragalactic Database (NED), ned.ipac.caltech.edu (exception: NGC 5128 has unspecified error, we take it as 10\%); For Col. (3): we use the method from \citet{laur10} (exceptions:  NGC 4261, NGC 4374, and NGC 4486 from \citet{marconi03} and NGC 5128 from \citet{cappe}); For Col. (4) The NED (exceptions: ARP 308, NGC 5532, and UGC 01841 have unspecified error, we take it as 10\%; NGC 1167 and NGC 4261 have unspecified error, we take it as 10\%; NGC 3801 and NGC 3862 have unspecified error, we take it as 10\%; NGC 
1167 and NGC 
4261 from \citet{nagar01} have unspecified error, we take it as 10\%; NGC 4651 have unspecified error, we take it as 10\%; UGC 02783 has unspecified error, we take it as 15\%. For CGCG 114-025, ESO 137-G006, IC 5063, NGC 5793, NGC 7075, UGC 11294, and WEIN 45 we estimate the core radio flux from the total one using the fitting formula from \citet{giovannini01}).}
\end{threeparttable}
\label{29table}
\end{table}


\begin{thebibliography}{64}
\expandafter\ifx\csname natexlab\endcsname\relax\def\natexlab#1{#1}\fi

\bibitem[{Aab et al.(2013b)}]{auger13} Aab, A. et al. (Pierre Auger Collaboration) 2013{\natexlab{b}}, arXiv: astro-ph/1307.5059

\bibitem[{Abbasi et al.(2010)}]{data2} Abbasi, R. et al. (HiRes Collaboration) 2010, Phys. Rev. Lett., 104, id. 161101

\bibitem[{Acciari et~al.(2009)}]{acciari09} Acciari V. A. et al. (VERITAS, MAGIC, VLBA M87 and H.E.S.S. Collaborations) 2009, Science, 325: 444

\bibitem[{Achterberg {et~al.}(2001)Achterberg, Gallant, Kirk, \&  Guthmann}]{acht} Achterberg, A., Gallant, Y.~A., Kirk, J.~G., \& Guthmann, A.~W. 2001, MNRAS, 328, 393

\bibitem[{Allard et al.(2008)}]{2008JCAP...10..033A} Allard, D., Busca, N.~G., Decerprit, G., Olinto, A.~V., \& Parizot, E.\ 2008, JCAP, 10, 33

\bibitem[{Abraham et al.(2007)}]{auger07} Abraham, J. et al. (Pierre Auger Collaboration) 2007, Science, 318, 938

\bibitem[{Abraham et al.(2008{\natexlab{a}})}]{auger08} Abraham, J. et al. (Pierre Auger Collaboration) 2008{\natexlab{a}}, Astropart. Phys., 29, 188

\bibitem[{Abraham et al.(2008{\natexlab{b}})}]{spectrum} Abraham, J. et al. (Pierre Auger Collaboration) 2008{\natexlab{b}}, Phys. Rev. Lett., 101, id. 061101

\bibitem[{Abraham et al.(2010{\natexlab{a}})}]{2010PhLB..685..239A} Abraham, J. et al. (Pierre Auger Collaboration) 2010{\natexlab{a}}, Phys. Lett. B, 685, 239

\bibitem[{Abraham et al.(2010{\natexlab{b}})}]{data1} Abraham, J. et al. (Pierre Auger Collaboration) 2010{\natexlab{b}}, Phys. Rev. Lett., 104, id. 091101

\bibitem[{Abreu et al.(2010{\natexlab{c}})}]{data1b} Abreu, P. et al. (Pierre Auger Collaboration) 2010{\natexlab{c}}, Astropart. Phys., 34, 314

\bibitem[{Abreu et al.(2011)}]{2011arXiv1107.4809T} Abreu, P. et al. (Pierre Auger Collaboration) 2011, arXiv: astro-ph/1107.4809 

\bibitem[{Abreu et al.(2013a)}]{2013ApJ...762L..13P} Abreu, P. et al. (Pierre Auger Collaboration) 2013{\natexlab{a}}, ApJL, 762, L13

\bibitem[{Bardeen(1970)}]{bardeen70} Bardeen, J.~M. 1970, Nature, 226, 64

\bibitem[{Becker \& Biermann(2009)}]{pbb09} Becker, J.~K. \& Biermann, P.~L. 2009, APh, 31, 138

\bibitem[{Bednarz \& Ostrowski(1998)}]{bo98} Bednarz, J. \& Ostrowski, M. 1998, Phys. Rev. Lett., 80, 3911

\bibitem[{Berezinsky {et~al.}(2006)Berezinsky, Gazizov, \& Grigorieva}]{berezinsky06} Berezinsky, V., Gazizov, A., \& Grigorieva, S. 2006, Phys. Rev. D, 74, id. 043005

\bibitem[{Biermann et~al.(2008)}]{biermann08} Biermann, P.~L. et~al. 2008, Proc. for Origin, Mass, Composition and Acceleration Mechanisms of UHECRs (CRIS 2008), Malfa, Italy
  (arXiv: astro-ph/0811.1848)

\bibitem[{Biermann \& Strittmatter(1987)}]{bs} Biermann, P.~L. \& Strittmatter, P.~A. 1987, ApJ, 322, 643

\bibitem[{Biretta {et~al.}(2002)Biretta, Junor, \& Livio}]{biretta02} Biretta, J.~A., Junor, W., \& Livio, M. 2002, NewAR, 46, 239

\bibitem[{Biretta {et~al.}(1999)Biretta, Sparks, \& Macchetto}]{biretta99} Biretta, J.~A., Sparks, W.~B., \& Macchetto, F. 1999, ApJ, 123, 351

\bibitem[{Bisnovatyi-Kogan \& Ruzmaikin(1976)}]{bisnovatyi-ruzmaikin} Bisnovatyi-Kogan, G.~S. \& Ruzmaikin, A.~A. 1976, Ap\&SS, 42, 401

\bibitem[{Blandford(1976)}]{bland76} Blandford, R.~D. 1976, MNRAS, 176, 465

\bibitem[{Blandford(2000)}]{blandford00} Blandford, R.~D. 2000, Physica Scripta, T85, 191

\bibitem[{Blandford \& K\"onigl(1979)}]{bk79} Blandford, R.~D. \& K\"onigl, A. 1979, ApJ, 232, 34

\bibitem[{Blandford \& Payne(1982)}]{bp82} Blandford, R.~D. \& Payne, D.~G. 1982, MNRAS, 199, 883

\bibitem[{Blandford \& Znajek(1977)}]{bz} Blandford R.~D. \&  Znajek R.~L. 1977, MNRAS, 179, 433

\bibitem[Blasi et al.(2000)]{blasi00} Blasi, P., Epstein, R., \& Olinto, A. V. 2000, ApJ, 533, L123

\bibitem[{Boldt \& Ghosh(1999)}]{boldt-ghosh} Boldt, E. \& Ghosh, P. 1999, MNRAS, 307, 491

\bibitem[Breitschwerdt et al.(1991)]{1991AA...245...79B} Breitschwerdt, D., McKenzie, J.~F., \& V\"olk, H.~J.\ 1991, A\&A, 245, 79 

\bibitem[{Bridle \& Perley(1984)}]{bridle&perley} Bridle, A. \& Perley, A. 1984, ARAA, 22, 319

\bibitem[{Brodatzki et al.(2011)}]{brodatzki11} Brodatzki, K. A., Pardy, D. J. S., Becker, J. K., \& Schlickeiser, R.

\bibitem[{Cappellari et~al.(2009)}]{cappe} Cappellari, M. et~al. 2009, MNRAS, 394, 660

\bibitem[{Caramete et~al.(2011a)}]{laur11a} Caramete, L.~I. et~al. 2011, arXiv: astro-ph/1106.5109

\bibitem[{Caramete \& Biermann (2011b)}]{laur11b} Caramete, L.~I. \& Biermann, P. L. 2011, arXiv: astro-ph/1107.2244

\bibitem[{Caramete \& Biermann (2010)}]{laur10} Caramete, L.~I. \& Biermann, P. L. 2010, A\&A, 521, 8

\bibitem[{Das et al.(2008)}]{2008ApJ...682...29D} Das, S., Kang, H., Ryu, D., \& Cho, J.\ 2008, ApJ, 682, 29 

\bibitem[Dolag et al.(2005)]{2005JCAP...01..009D} Dolag, K., Grasso, D., Springel, V., \& Tkachev, I.\ 2005, JCAP, 1, 9 

\bibitem[{Drenkhahn(2002)}]{drenkhahn} Drenkhahn, G. 2002, A\&A, 387, 714

\bibitem[{Du\c{t}an(2010)}]{eu10} Du\c{t}an, I., arXiv: astro-ph/1001:5434

\bibitem[{Du\c{t}an(2011)}]{eu11} Du\c{t}an, I., arXiv: astro-ph/1104.0825

\bibitem[{Du\c{t}an \& Biermann(2005)}]{eu04} Du\c{t}an, I. \& Biermann, P.~L. 2005, in Proc. of the 14th Course of the
  ISCRA, Erice, Italy, 2004. Eds. M. M. Shapiro, S. Todor, and J. P. Wefel. NATO science series II: Mathematics, physics and chemistry, vol. 209, p.175 

\bibitem[Everett et al.(2010)]{2010ApJ...711...13E} Everett, J.~E., Schiller, Q.~G., \& Zweibel, E.~G.\ 2010, ApJ, 711, 13

\bibitem[Everett et al.(2008)]{2008ApJ...674..258E} Everett, J.~E., Zweibel, E.~G., Benjamin, R.~A., et al.\ 2008, ApJ, 674, 258

\bibitem[{Falcke \& Biermann(1995)}]{fb95} Falcke, H. \& Biermann, P.~L. 1995, A\&A, 293, 665

\bibitem[{Falcke {et~al.}(1995)Falcke, Malkan, \& Biermann}]{fb95b} Falcke, H., Malkan, M.~A., \& Biermann, P.~L. 1995, A\&A, 298, 375

\bibitem[{Fang et al.(2012)}]{fang12} Fang, K., Kotera, K., \& Olinto, A. V. 2012, ApJ, 750, 16 

\bibitem[{Farrar et al.(2006)}]{farrar06} Farrar, G.~R., Berlind, A. A., \& Hogg, D. V. 2006, ApJ, 642, L89

\bibitem[{Farrar \& Gruzinov(2009)}]{Farrar09} Farrar, G.~R. \& Gruzinov, A. 2009, ApJ, 693, 329

\bibitem[{Fragile et al.(2012)}]{fragile12} Fragile, P.~C., Wilson, G., \& Rodriguez, R. 2012, MNRAS, 424, 524

\bibitem[{Gallant \& Achterberg(1999)}]{ga99} Gallant, Y.~A. \& Achterberg, A. 1999, MNRAS, 305, L6

\bibitem[{Giannios(2010)}]{giannios10} Giannios, D. 2010, MNRAS, 408, L46

\bibitem[{Giovannini et al.(2001)}]{giovannini01} Giovannini, G., Cotton, W. D., Feretti, L., Lara, L., \& Venturi, T. 2001, ApJ, 552, 508

\bibitem[{Greisen(1966)}]{greisen} Greisen, K. 1966, Phys. Rev. Lett., 16, 748

\bibitem[{Hada et al.(2011)}]{hada11} Hada, K. et al. 2011, Nature, 477 , 185

\bibitem[{Heinz \& Sunyaev(2003)}]{heinz-sunyaev} Heinz, S. \& Sunyaev, R.~A. 2003, MNRAS, 343, L59

\bibitem[{Hillas(1984)}]{hillas84} Hillas, A. M. 1984, ARA\&A, 22, 425

\bibitem[{Kaiser(2006)}]{kaiser} Kaiser, C.~R. 2006, MNRAS, 367, 1083

\bibitem[Kampert et al.(2013)]{2013APh....42...41K} Kampert, K.-H., Kulbartz, J., Maccione, L., et al.\ 2013, Astroparticle Physics, 42, 41

\bibitem[{Keshet \& Waxman(2005)}]{kw05} Keshet, U. \& Waxman, E. 2005, Phys. Rev. Lett., 94, id. 111102

\bibitem[{Kirk {et~al.}(2000)Kirk, Guthmann, Gallant, \& Achterberg}]{kirk00} Kirk, J.~G., Guthmann, A.~W., Gallant, Y.~A., \& Achterberg, A. 2000, ApJ, 542, 235

\bibitem[{Koide  {et~al.}(1999)}]{shinjiet99} Koide S.,  Shibata K.,  \&  Kudoh T. 1999, ApJ, 522, 727

\bibitem[{Lee et~al.(2008)}]{lee} Lee, S.-S. et~al. 2008, ApJ, 136, 159

\bibitem[{Longair(1994)}]{longair} Longair, M. 1994, High Energy Astrophysics, Vol.~2 (Cambridge Uni. Press), 259

\bibitem[{Lovelace(1976)}]{lovelace} Lovelace, R. V.~E. 1976, Nature, 262, 649

\bibitem[{Macchetto et~al.(1997)}]{macchetto} Macchetto, F. et~al. 1997, ApJ, 489, 579

\bibitem[{Macri et~al.(1999)}]{macri} Macri, L.~M. et~al. 1999, ApJ, 521, 155

\bibitem[{Marconi \& Hunt (2003)}]{marconi03} Marconi, A. \& Hunt, L. K. 2003, ApJ, 589, 21

\bibitem[{Markoff {et~al.}(2001)Markoff, Falcke, \& Fender}]{markoff01} Markoff, S., Falcke, H., \& Fender, R. 2001, A\&A, 372, L25

\bibitem[{Markoff {et~al.}(2005)Markoff, Nowak, \& Wilms}]{markoff05} Markoff, S., Nowak, M.~A., \& Wilms, J. 2005, ApJ, 635, 1203

\bibitem[{Marscher et~al.(2008)}]{marscher} Marscher, A.~P. et~al. 2008, Nature, 452, 966

\bibitem[{McKinney(2006)}]{mckinney06} McKinney J. C. 2006, MNRAS, 368, 1561

\bibitem[{McKinney \& Blandford (2009)}]{mckinney09} McKinney J. C. \& Blandford R. D. 2009, MNRAS, 394, L126

\bibitem[{Medina~Tanco et~al.(1998)}]{medina98} Medina~Tanco, G.~A., de~Gouveia~dal Pino, E.~M., \& Horvath, J.~E. 1998, ApJ, 492, 200

\bibitem[{Meisenheimer et~al.(2007)}]{meisenheimer} Meisenheimer, K. et~al. 2007, A\&A, 471, 453

\bibitem[{Meli {et~al.}(2008)Meli, Becker, \& Quenby}]{meli} Meli, A., Becker, J.~K., \& Quenby, J.~J. 2008, A\&A, 492, 323

\bibitem[{Mizuno {et~al.}(2007)}]{mizuno07} Mizuno Y.,  Nishikawa N.,  Koide S.,  Hardee P., \&  Fishman G.~J.  2007, in Proc. of the VI Microquasar Workshop: Microquasars and Beyond, Como, Italy.  Ed. B. Tomasso (Trieste: SISSA), p.45
  
  \bibitem[{Mizuno {et~al.}(2004)}]{mizuno04} Mizuno Y., Yamada, S., Koide, S., \& Shibata, K. 2004, 615, 389

  \bibitem[{Moskalenko {et~al.}(2009)Moskalenko, Stawarz, Porter, \& Cheung}]{moskalenko09} Moskalenko, I. V., Stawarz, L., Porter, T. A., \& Cheung, C. C. 2009, ApJ, 693, 1261

\bibitem[{Nagar {et~al.}(2001a)}]{nagar} Nagar, N.~M., Wilson, A.~S., \& Falcke, H. 2001, ApJ, 559, L87

\bibitem[{Nagar {et~al.}(2001b)}]{nagar01} Nagar, N. M., Falcke, H., \& Wilson, A. S. 2005, A\&A 435, 521

\bibitem[{Nakamura \& Asada(2013)}]{nakamura13} Nakamura, M. \& Asada, K. 2013, ApJ, 775, 11

\bibitem[{Niemiec \& Ostrowski(2006)}]{jn06} Niemiec, J. \& Ostrowski, M. 2006, ApJ, 641, 984

\bibitem[{Nishikawa {et al.}(2005)}]{ken05} Nishikawa K.-I.,  Richardson G.,  Koide S.,  Shibata K.,  Kudoh T.,  Hardee P., \& Fishman G.~J. 2005, ApJ, 625, 60

\bibitem[{Novikov \& Thorne(1973)}]{nt359} Novikov, I.~D. \& Thorne, K.~S. 1973, in Black Holes Les Astres Occlus, Ed. DeWitt, C. and DeWitt, B. S. (New-York: Gordon \& Breach), 359

\bibitem[{Parker(1958)}]{parker} Parker, R.~D. 1958, ApJ, 128, 664

\bibitem[{Protheroe \& Biermann(1996)}]{protheroe96} Protheroe, R.~J. \& Biermann, P.~L. 1996, Astropart. Phys., 6, 45

\bibitem[{Rybicki \& Lightman(1979)}]{ryb+ligh} Rybicki, G.~B. \& Lightman, A.~P. 1979, Radiative Processes in Astrophysics (John Wiley \& Sons)

\bibitem[Ryu et al.(1998)]{1998AA...335...19R} Ryu, D., Kang, H., \& Biermann, P.~L.\ 1998, A\&A, 335, 19

\bibitem[Ryu et al.(2008)]{2008Sci...320..909R} Ryu, D., Kang, H., Cho, J., \& Das, S.\ 2008, Science, 320, 909

\bibitem[Saba et al.(2013)]{saba13} Saba, I., Becker Tjus, J. K., \& Halzen, F. \ 2013, Astropart. Phys., 48, 30

\bibitem[{Sanders(1983)}]{sanders} Sanders, R.~H. 1983, ApJ, 266, 73

\bibitem[Sigl et al.(2003)]{2003PhRvD..68d3002S} Sigl, G., Miniati, F., \& Ensslin, T.~A.\ 2003, Phys. Rev. D, 68, 043002

\bibitem[{Slee et~al.(1994)}]{slee} Slee, O.~B. et~al. 1994, MNRAS, 269, 928

\bibitem[{Stanev(2010{\natexlab{a}})}]{stanev10b} Stanev, T. 2010{\natexlab{a}}, Review presented at the 2010 Vulcano workshop, Vulcano, Italy (arXiv: astro-ph/1011.1872)

\bibitem[{Stanev(2010{\natexlab{b}})}]{stanev10} Stanev, T. 2010{\natexlab{b}}, High Energy Cosmic Rays (Praxis Publishing Ltd, Chichester, UK, 2nd edition (First edition published in 2004.))

\bibitem[{Thorne(1974)}]{t74} Thorne, K.~S. 1974, ApJ, 191, 507

\bibitem[{Tingay et~al.(1998)}]{tingay} Tingay, S.~J. et~al. 1998, AJ, 115, 960

\bibitem[{van Velzen \& Falcke(2013)}]{vanVelzen2013} van Velzen, S. \& Falcke, H., The Intriguing Life of Massive Galaxies, Proceedings of the International Astronomical Union, IAU Symposium, Volume 295, pp. 271-271

\bibitem[{van Velzen et~al.(2012) van Velzen, Falcke, Schellart, Nierstenh\"ofer, \& Kampert}]{vanVelzen2012} van Velzen, S., Falcke, H., Schellart, P., Schellart, N., \& Kampert, K.-H. 2012, A\&A, 544, 18

\bibitem[{V\'eron-Cetty \& V\'eron (2006)}]{veron} V\'eron-Cetty, M. P. \& V\'eron, P. 2006, A\&A 455, 773

\bibitem[Vietri (1995)]{vietri95} Vietri, M. 1995, ApJ, 453, 883

\bibitem[{Vila \& Romero(2010)}]{vila10} Vila, G.~S. \& Romero, G.~E. 2010, MNRAS, 403, 1457

\bibitem[Waxman (1995)]{waxman95} Waxman, E 1995, Phys. Rev. Lett., 75, 386

\bibitem[Waxman \& Loeb (2009)]{waxman09} Waxman, E. \& Loeb, A. 2009, JCAP, issue 8, id. 026

\bibitem[Whysong \& Antonucci(2003)]{why-antonucci} Whysong, D. \& Antonucci, R. 2003, NewAR, 47, 219

\bibitem[{Zatsepin \& Kuzmin(1966)}]{zatsepin-kuzmin} Zatsepin, G.~T. \& Kuzmin, V.~A. 1966, Pis'ma Zh. Eksp. Teor. Fiz., 4, 114, [English translation in JETP Lett., 4, 78 (1966)]

\bibitem[{Zaw {et~al.}(2009)Zaw, Farrar, \& Greene}]{zaw09} Zaw, I., Farrar, G.~R., \& Greene, J.~E. 2009, ApJ, 696, 1218

\bibitem[{Znajek(1978)}]{znajek78} Znajek, R.~L. 1978, MNRAS, 185, 833

\end{thebibliography}
\end{document}